\input harvmac
\input epsf
%%%%%%%%%%%%%  DEFINITIONS %%%%%%%%%%%%%%%%%%%%%%%%%%%%%%%%%%%
\def\figin{\epsfcheck\figin}\def\figins{\epsfcheck\figins}
\def\epsfcheck{\ifx\epsfbox\UnDeFiNeD
\message{(NO epsf.tex, FIGURES WILL BE IGNORED)}
\gdef\figin##1{\vskip2in}\gdef\figins##1{\hskip.5in}% blank space instead
\else\message{(FIGURES WILL BE INCLUDED)}%
\gdef\figin##1{##1}\gdef\figins##1{##1}\fi}
\def\DefWarn#1{}
\def\figinsert{\goodbreak\topinsert}
\def\ifig#1#2#3#4{\DefWarn#1\xdef#1{fig.~\the\figno}
\writedef{#1\leftbracket fig.\noexpand~\the\figno}%
\figinsert\figin{\centerline{\epsfxsize=#3mm \epsfbox{#2}}}
\bigskip\medskip\centerline{\vbox{\baselineskip12pt
\advance\hsize by -1truein\noindent\footnotefont{\sl Fig.~\the\figno:}\sl\
#4}}
\bigskip\endinsert\noindent\global\advance\figno by1}
\def\ndt{{\noindent}}

\lref\NV{A.~Neitzke and C.~Vafa,
``${\cal N} = 2$ strings and the twistorial Calabi-Yau,''
{\tt hep-th/0402128}.
%%CITATION = HEP-TH 0402128;%%
}
\lref\witr{E. Witten,
``Perturbative gauge theory as a string theory in twistor space,''
{\tt hep-th/0312171}.
%%CITATION = HEP-TH 0312171;%%
}
\lref\orv{A. Okounkov, N. Reshetikhin, and C. Vafa, ``Quantum Calabi-Yau and
classical crystals'', {\tt hep-th/0309208}.}
\lref\inov{
A.~Iqbal, N.~Nekrasov, A.~Okounkov, and C.~Vafa,
``Quantum foam and topological strings,''
{\tt hep-th/0312022}.
%%CITATION = HEP-TH 0111068;%%
}
\lref\minet{M. Aganagic, R. Dijkgraaf, A. Klemm, M. Marino, and
C. Vafa, ``Topological strings and integrable hierarchies,''
{\tt hep-th/0312085}.}
\lref\don{S.K. Donaldson and R. Friedman, ``Connected sums of self dual
manifolds and deformations of singular spaces,'' Nonlinearity {\bf 2}, 197
(1989).}
\lref\dt{S. Donaldson and R. Thomas, ``Gauge theory in higher
dimensions,'' in {\it The geometric universe: science, geometry
and the work of Roger Penrose}, S. Huggett et. al. eds., Oxford
Univ. Press, 1998.}
\lref\pandet{D. Maulik, N. Nekrasov, A. Okounkov, and R. Pandharipande,
``Gromov-Witten theory and Donaldson-Thomas theory,''
{\tt math.AG/0312059}.}
\lref\gova{
R.~Gopakumar and C.~Vafa,
``M-theory and topological strings. I,''
{\tt hep-th/9809187};
%%CITATION = HEP-TH 9809187;%%
``M-theory and topological strings. II,''
{\tt hep-th/9812127}.
%%CITATION = HEP-TH 9812127;%%
}
\lref\asl{A.~S.~Losev, ``Perspectives of string theory,''
talk at the ``String theory at Greater Paris'' seminar, 2001.}
\lref\cv{C. Vafa, work in progress.}
\lref\ahs{M.F. Atiyah, N.J. Hitchin, and I.M. Singer, ``Self-duality
in four dimensional Riemannian geometry,'' Proc. Roy. Soc. London Ser. A
{\bf 362}, 425 (1978).}
\lref\taub{C.H. Taubes, ``The Existence of anti-self-dual conformal
structures,'' J. Diff. Geom. {\bf 36}, 163 (1992).}
\lref\bcov{M. Bershadsky, S. Cecotti, H. Ooguri, and C. Vafa, ``Kodaira-Spencer
theory of gravity and exact results for quantum string amplitudes,''
Commun. Math. Phys. {\bf 165} (1994) 311-428, {\tt hep-th/9309140}.}
\lref\naret{I. Antoniadis, E. Gava, K. S. Narain, and
 T. R. Taylor, ``topological
amplitudes in string theory,''
Nucl. Phys. {\bf B413} (1994) 162-184, {\tt hep-th/9307158}.}
\lref\dew{
G.~Lopes Cardoso, B.~de Wit, and T.~Mohaupt,
``Deviations from the area law for supersymmetric black holes,''
Fortsch.\ Phys.\  {\bf 48}, 49 (2000),
{\tt hep-th/9904005}.
%%CITATION = HEP-TH 9904005;%%
}
\lref\osv{H. Ooguri, A. Strominger, and C. Vafa, work to appear.}
\lref\vaug{C. Vafa, ``Superstrings and topological strings at
large $N$,''
J.\ Math.\ Phys.\  {\bf 42}, 2798 (2001),
{\tt hep-th/0008142}.
%%CITATION = HEP-TH 0008142;%%
}
\lref\kachet{S.~Kachru, M.~B.~Schulz, P.~K.~Tripathy, and S.~P.~Trivedi,
``New supersymmetric string compactifications,''
JHEP {\bf 0303}, 061 (2003),
{\tt hep-th/0211182}.
%%CITATION = HEP-TH 0211182;%%
}
\lref\louisone{
S.~Gurrieri, J.~Louis, A.~Micu, and D.~Waldram,
``Mirror symmetry in generalized Calabi-Yau compactifications,''
Nucl.\ Phys.\ B {\bf 654}, 61 (2003),
{\tt hep-th/0211102}.
%%CITATION = HEP-TH 0211102;%%
}
\lref\louistwo{
T.W. Grimm and J. Louis, ``The effective action of
${\cal N}=1$ Calabi-Yau orientifolds,''
{\tt hep-th/0403067}.
}
\lref\gvw{
S.~Gukov, C.~Vafa, and E.~Witten,
``CFT's from Calabi-Yau four-folds,''
Nucl.\ Phys.\ B {\bf 584}, 69 (2000)
[Erratum-ibid.\ B {\bf 608}, 477 (2001)],
{\tt hep-th/9906070}.
%%CITATION = HEP-TH 9906070;%%
}
\lref\tv{
T.~R.~Taylor and C.~Vafa,
``RR flux on Calabi-Yau and partial supersymmetry breaking,''
Phys.\ Lett.\ B {\bf 474}, 130 (2000),
{\tt hep-th/9912152}.
%%CITATION = HEP-TH 9912152;%%
}
%\StromingerNS
\lref\StromingerNS{
A.~Strominger,
%``Vacuum Topology And Incoherence In Quantum Gravity,''
Phys.\ Rev.\ Lett.\  {\bf 52}, 1733 (1984).
%%CITATION = PRLTA,52,1733;%%
}
\lref\withcs{E. Witten, ``Chern-Simons gauge theory as a
string theory,'' Prog. Math. {\bf 133}, 637 (1995),
{\tt hep-th/9207094}.}

\lref\lmn{A.~Losev, A.~Marshakov, and N.~Nekrasov,
``Small instantons, little strings and free
fermions,'' {\tt hep-th/0302191},
in: {\it From fields to strings: circumnavigating theoretical physics,}
Ian Kogan Memorial Volume,
M. Shifman, A. Vainshtein, and J. Wheater eds., World
Scientific, Singapore.}

%%%% Nikita's defs %%%%%%%%
%%%%%
\def\boxit#1{\vbox{\hrule\hbox{\vrule\kern8pt
\vbox{\hbox{\kern8pt}\hbox{\vbox{#1}}\hbox{\kern8pt}}
\kern8pt\vrule}\hrule}}
\def\mathboxit#1{\vbox{\hrule\hbox{\vrule\kern8pt\vbox{\kern8pt
\hbox{$\displaystyle #1$}\kern8pt}\kern8pt\vrule}\hrule}}

%%%%%%%%
%%%%%%%%%%%% Greek %%%%%%%%%%%%

\def\g{{\gamma}}
\def\e{{\epsilon}}

\def\u{{\Upsilon}}
\def\l{{\lambda}}

%%%%%
%%%%%%%Russian fonts%%%%%%%5
\chardef\tempcat=\the\catcode`\@ \catcode`\@=11
\def\cyracc{\def\u##1{\if \i##1\accent"24 i%
    \else \accent"24 ##1\fi }}
\newfam\cyrfam
\font\tencyr=wncyr10
\def\cyr{\fam\cyrfam\tencyr\cyracc}
%%%%%%%%%%%%%%%% Calligraphic letters  %%%%%%%%%%%%%

\def\bC{{\bf C}}

\def\bR{{\bf R}}

%%%%%%%%%%%% Derivatives  %%%%%%%%%%%
\def\p{\partial}

%%%%%%%%%%% letters with bar %%%%%%%%

%%%%%%%%%% Math symbols %%%%%%%%%%%%%

\def\Tr{{\rm Tr}}

\def\Det{{\rm Det}}

%%%%%%%%%%%% end of Nikita's defs %%%%%%%%%%%%%%%
\noblackbox
\newcount\figno
 \figno=1
 \def\fig#1#2#3{
 \par\begingroup\parindent=0pt\leftskip=1cm\rightskip=1cm\parindent=0pt
 \baselineskip=11pt
 \global\advance\figno by 1
 \midinsert
 \epsfxsize=#3
 \centerline{\epsfbox{#2}}
 \vskip 12pt
 {\bf Fig.\ \the\figno: } #1\par
 \endinsert\endgroup\par
 }
 \def\figlabel#1{\xdef#1{\the\figno}}
 \def\encadremath#1{\vbox{\hrule\hbox{\vrule\kern8pt\vbox{\kern8pt
 \hbox{$\displaystyle #1$}\kern8pt}
 \kern8pt\vrule}\hrule}}
\Title
{\vbox{
 \baselineskip12pt
\hbox{hep-th/0403167}
\hbox{CALT-68-2479}
\hbox{HUTP-04/A011}
\hbox{IHES/P/04/09}
\hbox{ITEP-TH-62/03}
}}
{\vbox{
 \centerline{S-duality and Topological Strings}
 }}
\centerline{ Nikita Nekrasov,$^{a}$\footnote{$^{\dagger}$}{On leave of
absence from: ITEP, Moscow, 117259, Russia}
Hirosi Ooguri,$^{b}$
and Cumrun Vafa$^{c}$}
\bigskip

\centerline{}
\centerline{$^{a}$ Institut des Hautes Etudes Scientifiques}
\centerline{Bures-sur-Yvette, F-91440, France}
\centerline{}
%\smallskip
\centerline{$^{b}$ California
Institute of Technology, Pasadena, CA 91125, USA}
\centerline{}
\centerline{$^{c}$ Jefferson Physical Laboratory, Harvard University}
\centerline{Cambridge, MA 02138, USA}
\bigskip
\bigskip
%\vskip .1in
\centerline{\bf Abstract}

In this paper we show how S-duality of type IIB superstrings
leads to an S-duality relating A and B model topological
strings on the same Calabi-Yau as
had been conjectured recently:  D-instantons
of the B-model correspond to A-model perturbative
amplitudes and D-instantons of the A-model
capture perturbative B-model amplitudes.
 Moreover
this confirms the existence of new branes in the two models.
 As an application we explain the recent results concerning  A-model
topological strings on Calabi-Yau
and its equivalence to the statistical mechanical
model of melting crystal.

%\smallskip
\Date{March 2004}

\newsec{Introduction}

Topological strings come in many types, of which A and B models
have been most thoroughly investigated.
In the perturbative A model version K\"ahler geometry of the Calabi-Yau
is probed while in the B model version the complex geometry
of the Calabi-Yau is captured.  On the other hand there is
an intriguing role reversal when it comes to the D-branes:
The D-branes of A-model correspond to Lagrangian
submanifolds, which naturally couple to the holomorphic 3-form,
whereas the D-branes of the B-model correspond to holomorphic
submanifolds, which naturally couple to K\"ahler structure.

In a recent paper \NV\ it was conjectured that this is not
accidental and that there is an S-duality relating the A and
B model topological strings on the same Calabi-Yau, where
D-brane instanton amplitudes of the B-model are computed by
worldsheet instantons of the A-model and non-perturbative
A-model amplitudes correspond to the B-model perturbative amplitudes.
This duality in particular predicts the existence of additional
branes called ``NS 5-brane'' in the A-model and ``NS 2-brane''
in the B-model.
This duality conjecture was motivated by explaining the equivalence
of the B-model topological string on the twistorial
Calabi-Yau  ${\bf CP}^{3|4}$ \witr\ with the S-dual
A-model topological string
on the same Calabi-Yau.  The two were conjectured to be
related by a Montonen-Olive S-duality \NV .

On the other hand a while back it was shown in \gova\ that the
D1 and D(-1) brane instantons of the B-model are counted by the A-model
worldsheet instantons.   More
recently
a description of A-model on Calabi-Yau
as the statistical mechanical model of melting
crystal was discovered in \orv .   This was also
interpreted as relating the A-model worldsheet instantons
to D1 and D(-1)-branes which would make more sense in a B-model
context \refs{\inov ,\pandet}. Also,  B model-like gauge-theoretic
calculations were
mapped to the A model worldsheet calculations with the topological gravity
observables turned on in \lmn. If there is an S-duality
between A and B models on the same manifold these statements
could be expected to be a consequence.

In this paper we explain how the S-duality of type IIB
superstring in ten dimensions {\it implies} the S-duality
between A and B model topological strings on the same
manifold.\foot{To be precise, the S-duality of type IIB superstrings
leads to the statement that D-brane instantons of B-model
are captured by A-model worldsheet instantons.  The reverse
statement follows by mirror symmetry.}  Moreover we
show how the results of \refs{\inov, \pandet}\ may be
viewed as an application of this S-duality.
In particular the S-dual of D(-1) and D1 brane correspond
to removing ``atoms'' or ``edges'' from the
 Calabi-Yau crystal introduced in
\orv .
%In particular the S-dual
%of D(-1) brane and D1 branes correspond to
They are gravitational
quantum foam in the A-model setup and can be interpreted
as blowing up the A-model geometry along the branes.

It is natural to consider applications of these ideas
to the twistorial Calabi-Yau in the absence
of any extra branes, where we consider the pure gravitational
theory: In particular the S-dual of D(-1) and D1 branes of the B-model
on ${\bf CP}^{3|4}$ at B-model strong coupling are
quantum foam in the twistor space ${\bf CP}^{3|4}$
in the weakly coupled A-model \cv \foot{
This suggests that D(-1) brane instantons should also play a role
in the twistorial Calabi-Yau proposed in \witr .}.
Moreover in the B-model setting the D1 branes can be viewed
as deforming the complex structure of twistor space
\minet\
which according to the results of \ahs\ can be mapped
back to a quantum foam in ${\bR}^4$.  For example the
D1 brane instanton wrapping a ${\bf P}^1$ cycle
of the twistor space, gets mapped to blowing up
a point in ${\bR}^4$ \refs{\taub ,\don}.  These are quite
exciting as they would lead to a stringy description
of quantum foam for ${\cal N}=4$ conformal supergravity!
In fact some aspects of quantum foam for ordinary
conformal gravity has been studied a long time ago \StromingerNS .

The organization of this paper is as follows:  In section 2
we review some basic facts about S-duality for type IIB
superstring.  In section 3 we show how this leads
to an S-duality for topological strings mapping
D-instantons of the B-model to worldsheet instantons
of the A-model.  In section
4 we consider an application of this idea and show how
this leads to the picture proposed in \refs{\inov ,\pandet}\
for computing A-model amplitudes in terms of D-instantons.
In section 5 we complete the discussion of S-duality by
extending it to the mirror statement.

\newsec{S-duality for type IIB superstrings}

In this section we briefly recall certain aspects
of S-duality for type IIB superstrings in ten dimensions
which is relevant
for us.  We will be concentrating on the ${\bf Z}_2\subset SL(2,{\bf Z})$
subgroup of the S-duality group corresponding to strong/weak
string coupling exchange.  We will
call the two dual theories by $B$ and $A$
(note that $A$ does not refer to type IIA superstring, but
to a dual type IIB superstring--
the reason for the choice of the letter $A$ becomes clear
when we talk about topological strings in the next section).

Let us denote the superstring coupling constant by $g_B$.
The strong/weak duality relates this to a dual type IIB
coupling constant
which we denote by $g_A$:
\eqn\cdu{g_A={1\over g_B}}
As usual the coupling can be complexified, but we will
deal with it in this way because all the relevant expressions
are analytic.  The metrics of the two theories are related
by
\eqn\medu{{g^B_{\mu \nu}\over g_B}=g^A_{\mu \nu}.}
We will be interested in Calabi-Yau 3-fold backgrounds.  It
is natural to ask how the K\"ahler form $k$ and the holomorphic
3-form $\Omega$ of the Calabi-Yau transform under S-duality.  By $\Omega$
we mean the normalized holomorphic three form such that
$\Omega \wedge {\overline \Omega_0}$ gives the volume form.
Here we vary the holomorphic form $\Omega$ while keeping
${\overline \Omega_0}$ fixed and so all the volume
dependence in captured by the holomorphic part $\Omega$
($i.e.$ we can view $\Omega$
as mirror to the K\"ahler volume $k\wedge k \wedge k$). {}From \medu\ it follows that
\eqn\kdu{{k_B\over g_B}=k_A}
\eqn\odu{{\Omega_B \over {g_B^{3}}}=\Omega_A}

Under S-duality we have the following exchange of the branes
$$D1\leftrightarrow F1$$
$$D5\leftrightarrow NS5$$
where $F1$ denotes the fundamental string.  It will be convenient
to discuss what holomorphic configurations
of $D1$ branes and $F1$ branes couple to.  For $D1$ brane
this is given by
$$\int_{D1} {k_B\over g_B}+i B_R$$
where the first term comes from the volume of $D1$ brane and
the second term denotes the fact that $D1$ brane is charged under
the RR 2-form field $B_R$.  Similarly holomorphic $F1$ couples to
$$\int_{F1}k_A+iB_{NS}$$
where $B_{NS}$ denotes the NS-NS 2-form field.  It is natural
to define the fields
$${\hat k}_B=k_B+i g_B B_R$$
$${\hat k}_A=k_A+i B_{NS}$$
in terms of which the holomorphic $D1$ brane couples to ${\hat k}_B/g_B$
and the holomorphic $F1$ branes couple to ${\hat k}_A$.  Under
S-duality we have the map
$${{\hat k}_B\over g_B}={\hat k}_A$$
reflecting the fact the $D1$ and $F1$ exchange under
S-duality.  This is what replaces \kdu\ when the B-fields
are turned on.  For simplicity of notation for the rest
of the paper we drop the hats
and denote the K\"ahler forms, including the $B$ fields by $k_B$
and $k_A$.

\newsec{Type IIB superstrings and A and B model topological strings}

Consider type IIB superstrings compactified on a Calabi-Yau threefold $M$.
Consider the topological strings on $M$.  There are two versions, $A$ and
$B$,
of topological strings on $M$ and they are known to compute
`F-terms' for type IIB superstrings:
B-model topological strings compute F-terms for vector multiplets
\refs{\bcov , \naret}, and A-model topological strings
computes F-terms for hypermultiplets of type IIB superstrings \naret.
  For each worldsheet genus $h$, each topological
string computes  correction for a different, but
unique F-term.  Moreover the topological string amplitude
at given genus corresponds
to the  F-term correction  in the superstring coming
from the {\it same} genus  amplitude.  Thus in this
way {\it we can identify the topological string coupling
constant with the superstring coupling constant}, bearing
in mind that each term of the topological string computes
different amplitudes for the superstring.  In some
cases these different terms can be identified, as in the
context of black holes \refs{\dew ,\osv }.

Consider topological B-model.  We ask if there are any non-perturbative
corrections.
 Let us consider the ones coming
from D1 brane instantons.  These are corrections
to the hypermultiplet moduli.  As discussed
in the previous section these couple to $k_B/g_B$.
However, under S-duality D1 brane and F1 brane are exchanged, and
the corresponding instantons get mapped to holomorphic worldsheet
instantons.
  These are precisely the objects of relevance
for the A-model topological string!   Thus these D-brane instanton
corrections
are mapped to A-model perturbative worldsheet correction of the dual
theory.  We are thus led
to view the non-perturbative completion of B-model as including
the perturbative modes of the dual A-model.  Moreover the D1 brane instantons
of the B-model are captured by the dual worldsheet A-model instantons on
the same manifold.
This was conjectured in \NV\ and we now see
it can be inferred from superstring S-duality for type IIB
superstrings.  In particular we have
$$g_A=1/g_B$$
$$k_A=k_B/g_B$$

What is the interpretation of the various D-branes
in the dual A-model?  The interpretation of D5 brane dual
is an NS 5-brane.  This provides evidence for the existence of a new
brane in A-model, which was conjectured in \NV .  It
is also natural to expect that a holomorphic Chern-Simons theory lives
on this NS 5-brane, as would follow
from type IIB string duality.

The story for D1 branes and D(-1) brane instantons
are expected to be more tricky as we know from superstrings.
Consider in particular a contribution in the B-model involving
$N$ D(-1) brane instantons and a D1 brane wrapped
over a 2-cycle class $(c)$.  These should contribute
to the amplitude by a factor of
\eqn\fac{{\rm exp}[-N/g_B - k_B(C)/g_B]={\rm exp}[-Ng_A -k_A(C)]}
This factor for the A-model is not the usual one and in
particular the $g_A$ appears in a very different way
from the conventional form.   However
in fact this form of the contribution
anticipates a periodicity in $g_A$ by shifting
by $g_A \rightarrow g_A+ 2\pi i$ which was predicted
in \gova\ based on considerations
of embedding topological string in superstrings and
viewing it as counting wrapped D2 brane degeneracies.
Thus it means that D(-1) brane and D1 branes repackage
the A-model amplitudes in a different way.

In fact the story is much better:  This form of the
A-model expansion is deeply connected to the recent
discovery of the description of A-model topological
strings on Calabi-Yau as a melting crystal \orv\
reformulated as a $U(1)$ gauge theory \refs{\inov , \pandet}.
Moreover in that context the D(-1) brane and D1 brane instantons
did end up having a geometric meaning in the dual A-model context:
They correspond to quantum gravitational foam for the A-model,
where the space was blown up along the location of D(-1) brane points
and D1 brane curves of the size $g_A$.  Thus this gives
a more precise meaning as to the role of D(-1) and D1 brane
instantons viewed from the dual A-model.

So far we have only shown that S-duality predicts
a duality between D1 and D(-1)-brane instantons of the
B-model and perturbative amplitudes of the A-model, and
have made some qualitative checks.  Can we use this
duality to gain further insight into the A-model amplitudes?
In particular can we use the D-instanton sum of the B-model
as a new way to compute the A-model amplitudes?

In fact we will argue in the next section that this is possible and
that it has already been done \refs{\inov , \pandet}!

\newsec{Holomorphic Chern-Simons and A-model}

In the previous section we have argued that the A-model
amplitudes should be computable by summing up the
D(-1) and D1 brane instantons.  The natural question
is how to compute these instanton contributions directly.

There is one natural way this can be done.  We can consider
a single D5 brane wrapped over the Calabi-Yau.  In this context
the sum over the D(-1) and D1 brane instantons is the same
as the sum over the various sectors of the $U(1)$ bundle.
On the D5 brane, in the topological B-model context,
lives a $U(1)$ holomorphic Chern-Simons theory \withcs .  Moreover
the $N D(-1)$ branes and $D1$ brane
wrapped over a cycle $[C]$ get mapped to a gauge theory
configuration for holomorphic Chern-Simons having
$$ch_3=N$$
$$ch_2=[C].$$
In particular the ``$U(1)$'' theory is a stringy $U(1)$
theory which supports
such non-trivial configurations.
Of course holomorphic Chern-Simons will also
have perturbative contributions.  These
will not be relevant for the above instanton terms, if we wish
to compute A-model closed string amplitudes.  So
roughly speaking we wish to compute
$$Z_{hCS}/Z^{pert.}_{hCS},$$
and we are organizing the instanton sum, using \fac\ as
$$\exp\left[-g_A \int ch_3-\int k_A\wedge ch_2\right]$$
We thus expect to have a $U(1)$ gauge theory with
the above weight which is morally the non-perturbative
contributions of the holomorphic Chern-Simons.

There are two proposals for what this gauge theory may be \refs{\inov,
\pandet}.
The formulation in \inov\ involves a twisted version of maximally
supersymmetric
Yang-Mills on the Calabi-Yau.  The one in \pandet\ is morally equivalent
to holomorphic Chern-Simons theory on the Calabi-Yau (in the sense
that it ``counts'' holomorphic bundles defined mathematically in
\dt ).  It is not known if the two are equivalent.  Both of them
 localize to holomorphic bundles (or more precisely ideal
 sheaves) on the Calabi-Yau.

Let us discuss the one in \inov\ from the perspective
of the present paper:
If we think about the D5 brane in the superstring, it
supports the partially twisted maximally supersymmetric Yang-Mills theory.
Its instanton equations (for $N$ D5 branes) are given by:
\eqn\inste{\eqalign{ & F^{2,0}_{A} = {\bar\p}^{\dagger}_{A} {\varphi} \cr
& F^{0,2}_{A} = {\p}^{\dagger}_{A}{\bar\varphi} \cr
& F^{1,1}_{A} \wedge k^2  = [ {\varphi}, {\bar\varphi} ] + {\l} \cdot
k^3 \cr}}
where $\varphi$ is the adjoint-valued $(3,0)$-form on the worldvolume of the
brane, which is the twisted complex Higgs field of the six dimensional gauge
theory. The equations \inste\ depend on the choice of complex structure, and
K\"ahler structure. However, as it often happens with the twisted gauge
theories, small variations of the K\"ahler structure should  not affect gauge
theory correlation functions.

Moreover, on Calabi-Yau manifolds, on the solutions of \inste\ the
$(3,0)$-form
vanishes, $\varphi = 0$. Then the equations \inste\ can be reformulated as
$F^{0,2}_{A}=0$ and its conjugate, while the last equation together with
ordinary $U(N)$ gauge symmetry combine to the complexified gauge symmetry,
$GL(N, {\bC})$. But then we are discussing precisely the equations of motion
of the holomorphic Chern-Simons theory on the same manifold, modulo complex
gauge transformations, which is the gauge symmetry of hCS!

We should be more careful, though. In the physical D5 brane theory there are
four scalars,  of which two are twisted into ${\varphi}, {\bar\varphi}$,
while
the other two remain intact, ${\Phi}, {\bar\Phi}$. The contribution of the
instanton solution \inste\ to the gauge theory path integral is given by the
ratio of determinants:
\eqn\instdet{ {\Det_{{\Omega}^{0,0}}{Ad({\Phi})}
\Det_{{\Omega}^{0,2}}{Ad({\Phi})}
\over \Det_{{\Omega}^{0,1}}{Ad({\Phi})} \Det_{{\Omega}^{0,3}}{Ad({\Phi})}}}
where the numerator comes from Faddeev-Popov ghosts and the fermions which
couple to the equations \inste, while the denominator comes from the
fluctuations of the gauge fields $A$ and Higgs $\varphi$. Now, the
remarkable fact is that on Calabi-Yau manifolds the ratio of the
determinants \instdet\ is actually $\Phi$-independent, apart from
inessential universal perturbative piece \refs{\inov ,\pandet}. We present
some
details of this computation in the Appendix.
This makes it possible to identify
the instanton contributions with those of hCS theory, as we had
anticipated from the S-duality of topological strings.

\newsec{Completing the S-duality of A and B models}

So far we have argued that the non-perturbative amplitudes
of the B-model are captured by perturbative amplitudes
of the A-model.  To complete the story we need to argue,
as has been conjectured in \NV , that the non-perturbative
amplitudes of the A-model are computed by the perturbative
amplitudes of the B-model.  This is clearly plausible,
because the instantons of the A-model are D2 branes wrapped
over Lagrangian 3-cycles of CY. Moreover one expects
that the perturbative B-model ``counts'' D3 branes
wrapped over Lagrangian 3-cycle.
We will now present further arguments to support this.
In fact this follows from mirror symmetry.  If the non-perturbative
amplitude of B-model is computed by perturbative amplitudes
of the A-model, then the mirror statement is that the non-perturbative
amplitude of the A-model is computed by perturbative
amplitudes of the B-model.
Thus to argue this statement it is natural to consider
type IIA superstrings on a Calabi-Yau.

So consider type IIA superstring compactified on the
Calabi-Yau $M$.
To this end it is natural to promote the A-model
threeform $\Omega_A$ to
$${\hat \Omega}_A=\Omega_A + i g_A C_R$$
where $C_R$ is an RR 3-form field for type IIA superstrings.
The analog of S-dual of $C_R$ is a $NS$ field $C_{NS}$ whose
flux corresponds to non-integrability of the complex
structure.  This was anticipated by mirror symmetry \vaug\
and has been verified in a number of
examples \refs{\louisone, \kachet , \louistwo}.  The dual
three form is defined by
$${\hat \Omega}_B=\Omega_B+i g_B^2 C_{NS}$$
This is forced by the condition that $C_R\leftrightarrow C_{NS}$
under the topological S-duality, and the fact that \odu\
predicts
$${{\hat \Omega_B}\over g_B^2}\leftrightarrow  {{\hat \Omega_A}
\over g_A} \ . $$
Note that the fact that in the A-model
the worldsheet instantons can end on Lagrangian D-branes,
gets mapped by the S-duality to the statement that the D1 brane instantons
can end on Lagrangian NS branes, which was called ``NS 2-branes''
in \NV .  Their geometric meaning is that they correspond to a source
for lack of integrability of the complex structure of the Calabi-Yau
in the B-model.

Let us drop the hats from $\Omega$'s keeping
in mind that we can add these fields.
As further evidence for this duality, let us recall
the term considered in \NV :  It was argued
there that there is a term in the topological B-model given by
$$\int {\Omega_B \wedge dk_B\over g_B^2}$$
note that this includes the term
${i\over g_B}\int \Omega_B \wedge dB_R$.  Another way to
explain the existence of this term is to note that this
is a superpotential term generated by RR flux $H_R=dB_R$
\refs{\gvw, \tv }.  The S-dual of the above term, using
the above transformations is given by
$$\int{\Omega_A \wedge dk_A \over g_A^2}$$
which is also generated in the A-model as explained
in \NV .

That perturbative B-model can be reformulated as a sum over
Lagrangian D-branes is amusing and it would be interesting
to see if this leads to another computational scheme for the
 B-model.

\bigskip
\ndt {\bf Acknowledgments.}
\bigskip
\ndt
We would like to thank M. Aganagic, S. Gukov,
 A. Kapustin, S. Katz,
L. Motl, A. Neitzke, A. Okounkov, A. Strominger and C. Taubes for
valuable discussions.  In addition NN would like to thank
A. Losev\foot{In fact, A.~Losev
has been advocating the idea of S-dualities and even M-theories in the context
of topological strings for a while \asl .} for interesting
discussions.

The research of NN is partly supported by {\cyr RFFI} grant 03-02-17554
and
by the grant {\cyr NSh}-1999.2003.2 for scientific schools.
The research of HO is supported in part by DOE grant DE-FG03-92-ER40701.
The research
of CV is supported in part by NSF grants PHY-0244821 and DMS-0244464.
\appendix{A}{Instanton contribution}
We consider the instanton contribution to the partition function
of the maximally supersymmetric $U(N)$ gauge theory on ${\bR}^6$. In order to regularize the possible
infrared divergencies we shall work equivariantly with respect to the
rotations of ${\bR}^6$. More precisely, in order to preserve some fermionic
symmetry one should combine the rotations of ${\bR}^6$ with some R-symmetry
rotations. The simplest possibility is to compensate rotation
generated by the $SO(6)$ matrix
\eqn\sosm{{\Omega} = \pmatrix{ 0 & {\e}_1 & & & & \cr
-{\e}_1 & 0 & & & &\cr
& & 0 & {\e}_2 & & \cr
& & -{\e}_2 & 0 & & \cr
& & & & 0 & {\e}_3 \cr
& & & & -{\e}_3 & 0}}
by twisting one of the complex scalars ${\varphi}$ by $e^{- i
\left( {\e}_1 + {\e}_2 + {\e}_3 \right)}$. The remaining scalar, $\Phi$,
is the Higgs field and could potentially enter the instanton contributions,
thus spoiling the conjectured hCS/SYM duality.

In fact, the contribution of a given instanton is then given by:
\eqn\insttc{{\exp} \int_{0}^{\infty}
{dt \over t} {{{\Tr} e^{t {\Phi}} {\Tr} e^{-t{\Phi}}}
\over {(1- e^{t{\e}_1})(1- e^{t{\e}_2})(1-e^{t{\e}_3})}}
}
and
\eqn\insttct{\eqalign{&{\Tr} e^{t{\Phi}} = \sum_{l=1}^{N} e^{t a_l}
\left[ 1 - (1- e^{t{\e}_1})(1- e^{t{\e}_2})(1-e^{t{\e}_3})
\sum_{(i,j,k) \in {\pi}_l} e^{t ({\e}_1 ( i-1) + {\e}_2 ( j-1) + {\e}_3 (
k-1))} \right],\cr
&\cr}}
where ${\e}_1, {\e}_2, {\e}_3$ are the equivariant parameters of the
$SO(6)$ rotation. They can also be mapped to the components of the
field strength of the graviphoton field. The $U(N)$
instantons are labeled by the $N$-tuples of three dimensional partitions
${\pi}_l$, $l=1 , \ldots , N$ and the classical value of the Higgs field
${\Phi}$ is given by the diagnal matrix with entries
 $(a_1, \ldots, a_N)$. The extra terms in \insttct\
are the quantum corrections, which lead to the modified expressions
for the $\langle {\Tr} {\Phi}^k \rangle$ expectation values.

Now the crucial point, heavily exploited in \pandet\inov\
is that for the Calabi-Yau choice of the rotation
parameters, $i.e.$ for ${\e}_1 + {\e}_2 + {\e}_3 = 0$ the determinants
\insttc\ almost cancel, leaving only the overall sign $(-)^{\sum_l \vert
{\pi}_l \vert}$ and the universal
perturbative factor independent of $\pi_l$:
\eqn\instpert{\eqalign{& Z^{pert} = {\exp} \sum_{l, m} {\g} \left( {a_l - a_m \over
\Lambda} \right) \cr
& \cr
& {\g}(x) = {d\over ds}\Biggr\vert_{s=0} \left( {1\over {\Gamma} (s)}
\int_{0}^{\infty}
{dt\over t} t^s
{e^{- t x} \over{(1- e^{t{\e}_1})(1- e^{t{\e}_2})(1-e^{t{\e}_3})}
}
\right). \cr}}

\listrefs
\end